\documentclass[prl,preprint,showpacs,superscriptaddress]{revtex4}
\usepackage{wrapfig}
\usepackage{graphicx}
\usepackage{epstopdf}
\usepackage{gensymb}

\begin{document}
\title{Molecule-assisted ferromagnetic atomic chain formation}

\author{Manohar Kumar}
\altaffiliation{Current address: O.V. Lounasmaa Laboratory, Aalto University, Puumiehenkuja 2B Otaniemi, Espoo, Finland }
\affiliation{Huygens-Kamerlingh Onnes Laboratory, Leiden University, Niels Bohrweg 2, 2333 CA Leiden, The Netherlands}
\author{ Kiran Kumar Vudya Sethu}
\affiliation{Huygens-Kamerlingh Onnes Laboratory, Leiden University, Niels Bohrweg 2, 2333 CA Leiden, The Netherlands}
\author{Jan M. van Ruitenbeek}
\email[Corresponding author:  ]{ruitenbeek@physics.leidenuniv.nl}
\affiliation{Huygens-Kamerlingh Onnes Laboratory, Leiden University, Niels Bohrweg 2, 2333 CA Leiden, The Netherlands}


\begin{abstract}
One dimensional systems strongly enhance the quantum character of electron transport. Such systems can be realized in 5d transition metals Au, Pt and Ir, in the form of suspended monatomic chains between bulk leads. Atomic chains between ferromagnetic leads would open up many perspectives in the context of spin-dependent transport and spintronics, but the evidence suggests that for pure metals only the mentioned three 5d metals are susceptible to chain formation. It has been argued that the stability of atomic chains made up from ferromagnetic metals is compromised by the same exchange interaction that produces the local moments.  Here we demonstrate that magnetic atomic chains can be induced to form in break junctions under the influence of light molecules. Explicitly, we find deuterium assisted chain formation in the 3d ferromagnetic transition metals Fe and Ni. Chain lengths up to eight atoms are formed upon stretching the ferromagnetic atomic contact in deuterium atmosphere at cryogenic temperatures.  From differential conductance spectra vibronic states of D$_2$ can be identified, confirming the presence of deuterium in the atomic chains.
\end{abstract}

\maketitle
\section{Introduction}
The spontaneous formation of chains of metal atoms was observed experimentally for Au atomic contact breaking cycles by two groups simultaneously in 1998~\cite{Yanson1998,Ohnishi1998}. Since, atomic chain formation for other transition metal elements, in pure form or assisted by small molecules or impurities, has been reported by various groups~\cite{Cuveas1998, Rubio2003, Scheer2006, Bettini2006, Thijssen2006, Csonka2006,  Fischer2007, Halbriter2007, Smit2009, Makk2012}. The experiments were mainly performed using the mechanically controllable break junction (MCBJ) technique, or using scanning tunneling microscope (STM) methods.  In a typical break junction experiment, such as reported by Yanson \textit{et al.}, a wire of the metal of interest, prepared with a weak link in its midpoint, is stretched in a controlled fashion using a piezo-electric element, while the conductance of the sample is recorded simultaneously. During the breaking of the wire the weak link is reduced to atomic size just before it breaks. The controlled stretching in the last stages of the atomic contact can lead to pulling new atoms into a chain arrangement, depending on the type of metal under study. The process of atomic chain formation requires the breaking force $F_0$ to be large.  This force equals the maximum string tension in the atomic chain at the point of inflection of the total energy as a function of stretching. Further, the difference in cohesion energy of atom in the leads ($E_{Lead}$) and the chain ($E_{Chain}$), needs to be small enough $\delta E = E_{Chain}- E_{Lead}$. When an atomic contact is elongated, the energy of the system increases to the point where an atom from the lead can overcome the energy required to join the chain. Upon addition of this new atom in the chain, the interatomic bond length relaxes, thus lowering the mechanical strain energy of the system. 
\newline Apart from intrinsic atomic chain formation for pure systems, chain formation induced by small molecules was reported by Thijssen \textit{et al.}~\cite{Thijssen2006}. Depending on the type of metal, the presence of gases such as D$_2$ , H$_2$ , O$_2$ , N$_2$ may lead to a linear metal-molecule bond that is stronger than the pure linear metal-metal bond, which increases in the break force ($F_0$). This was proposed to explain the  observations by Thijssen \textit{et al.} of very long chain formation in Ag and Cu break junctions in the presence of O$_2$  molecules. Here, we report a similar approach to investigate chain formation in the ferromagnetic atomic contacts for Fe and Ni in presence of deuterium, D$_2$. Under clean conditions of cryogenic vacuum, ferromagnetic atomic contacts for Fe, Ni and Co fail to form the atomic chains of more than 2 atoms long \cite{Untiedt2004, Calvo2008, Calvo2009}. In contrast, in presence of D$_2$  ferromagnetic atomic contacts readily form stable and long atomic chains. Note that the use of D$_2$ in stead of H$_2$ is for practical purposes only. Indeed, similar effects of H$_2$ induced formation of short atomic chains have recently been reported for Co and Pd \cite{Nakazumi2010,Kiguchi2010}.

\subsection{Experimental procedure}
Ni and Fe atomic contacts are formed from metallic wires of purity $99.998\% $ by breaking under cryogenic vacuum, at liquid helium temperatures. We use the mechanically controllable break junction technique, similar to that used by Thijssen {\it et al.}~\cite{Thijssen2006}. The contact breaking process, induced by stretching the metallic wire at the weak spot, can be viewed as the formation of a neck than gradually becomes narrower. In the last stages of stretching an atomic size contact is formed, just before breaking. Stretching of the contact with sub-{\AA}ngstr{\" o}m precision is facilitated by the use of a piezo element. We start by characterizing the native ferromagnetic atomic junctions by collecting conductance breaking traces into a histogram of conductance values, and verifying the expected characteristics of atomic ferromagnetic junctions (see Supplementary Information~\cite{Suppl.Info.}).

 After characterization of the atomic contacts, deuterium gas (D$_2$, $99.999\%$) is introduced from a vapor source at room temperature through a capillary tube leading to the contact. This procedure requires three steps in order to reduce contamination of the atomic contacts with unspecified molecules. First, all the tubes connected top of the dipstick, which is sealed by a needle valve, are flushed with D$_2$ gas. Next, the tubes are pumped back to base pressure (about $10^{-6}$ mbar). The capillary tube, running from top of the dipstick to the cryogenic break junction stage at its bottom, ends in a small detachable nozzle with an opening facing towards the sample. The capillary tube and nozzle are baked out at $150^{\circ}$C for a day before cooling down the dipstick to liquid He temperature. By cooling to liquid helium temperatures residual gases condense to the walls, creating a very good (cryogenic) vacuum.  D$_2$ is introduced into the capillary by opening the needle valve, while keeping the nozzle at its base temperature (about 5K). During the process of introducing the gas the junction is continuously cycled between broken and closed states and conductance traces are continuously monitored. As soon as a molecule is caught in the atomic contact the atomic signature of the contact is suppressed and the molecular signature is seen in the conductance traces: for D$_2$, steps appear in the conductance traces near the quantum unit of conductance,  $1\rm{G}_0 = 2e^2/h$, resulting in a peak at  $1\rm{G}_{0}$ in the conductance histogram~\cite{Smit2002}. As soon as this molecular signature is seen the needle valve is closed and the tubes connected to it are pump down to the base pressure ($10^{-6}$mbar). All conductance measurements for this characterization procedure are two-point dc measurements at a bias of 50mV. For differential conductance measurements of individual stable configurations a lockin amplifier is used with a modulation signal of amplitude 2mV and frequency 2.777kHz, while sweeping dc  bias voltage from -80mV to 80mV~\cite{Manoharthesis2012}. The measurements of shot noise are done using the two-channel cross-spectrum measurement up to 100kHz, as described in Refs.~\cite{Brom1999,Djukic2006,Kumar1996,Manoharthesis2012}.

\subsection{Experimental observations}
\begin{figure}[h!]
\centering
\includegraphics[scale =0.55]{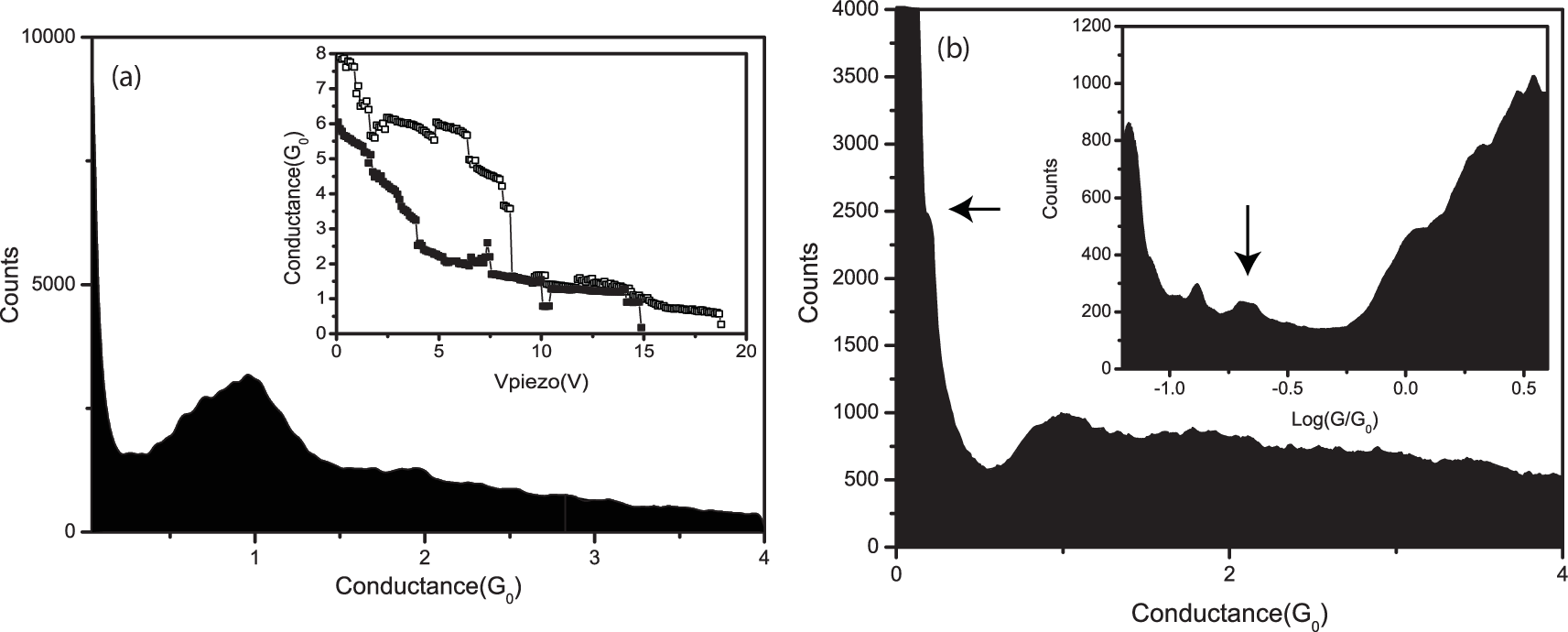}
\caption [Conductance histograms for Ni-D$_2$-Ni  and Fe-D$_2$-Fe atomic contacts]{Conductance histogram for Ni-D$_{2}$-Ni atomic contacts (a), constructed by collecting measured conductance points from $10,000$ conductance breaking traces. The inset shows two typical traces used in making this histogram. The conductance histogram for Fe-D$_2$-Fe atomic contacts in (b) is obtained from $5000$ conductance traces. The inset shows a logarithmic conductance histogram made from same traces, which shows the presence of a small peak at $0.25\rm{G}_{0}$.  \label{Gtrace_Ferro} }
\end{figure}

Ferromagnetic junctions are gently broken in the cryogenic environment and the conductance is recorded during the breaking process. The last plateau of the breaking trace signifies a junction cross section corresponding to a single atom. This last plateau is of our interest. In the cases of Ni and Fe ferromagnetic atomic contacts the last plateaus in the conductance traces are seen  near 1.3G$_{0}$ and 2.1G$_{0}$, respectively. The ferromagnetic atomic contacts tend to form short plateaus only, corresponding at best to a short chain of only two atoms connecting the leads at either side. Typical conductance traces and conductance histograms for Fe and Ni atomic contacts are shown in the supplementary information~\cite{Suppl.Info.}. Upon addition of D$_2$ the last conductance plateau is shifted to values near 1G$_{0}$. Two examples of conductance traces for Ni-D$_2$-Ni are shown in inset of Fig.~\ref{Gtrace_Ferro}a, plotted against the piezo voltage that controls the displacement of the two wire ends with respect to each other. A conductance histogram built from thousands of such traces is shown in the main panel of Fig.~\ref{Gtrace_Ferro}a. The conductance histogram shows a broad peak at 1G$_{0}$ with a decaying tail at higher conductance and a small shoulder peak at 2G$_{0}$. The strong peaks at 1.2G$_{0}$ and 1.6G$_{0}$ seen for pure Ni atomic contacts are completely suppressed in the presence of D$_2$. Translation of the piezo voltage scale to into a length scale, using a calibration procedure as described in ref.~\cite{Untiedt2002,Manohar2012}, shows that the length of the last plateau for this system is limited to about 3 atoms. 

\begin{figure}[t!]
\centering
\includegraphics[scale =0.40]{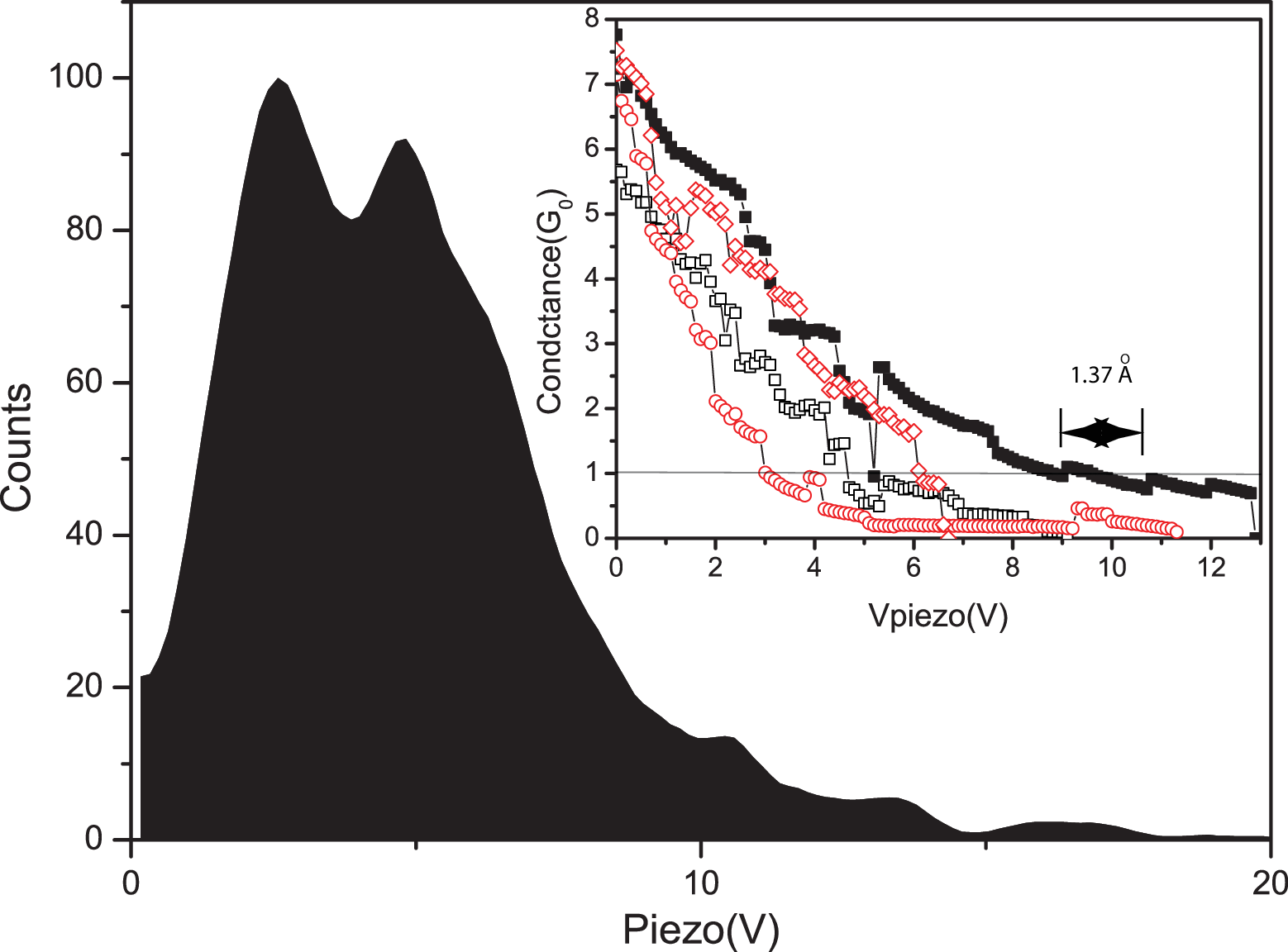}
\caption [Length histogram for Fe-D$_2$-Fe atomic contacts]{ Length histogram for Fe-D$_2$-Fe atomic contacts obtained from conductance traces, showing multiple peaks with first inter peak distance of 1.63\AA. The length histogram was recorded in  the conductance window of $1.5\rm{G}_{0}$ and $0.5\rm{G}_{0}$.  The inset shows typical conductance traces for Fe-D$_2$-Fe. The black trace shows the presence of $4$ steps of 1.37{\AA}  in the relevant window, which indicates the formation of a $5$ atom long chain. 
The other traces show the presence of extended structures at very low conductance of $0.2-0.3\rm{G}_{0}$, which are not included in this length histogram because they fall outside the selection window. The length axis of length histogram is shown in units of voltage on the piezo element. The appropriate proportionality constant is 1.45$\rm{V}/\rm{\AA}$, using the known inter peak distance in length histograms for Au atomic contacts for calibration~\cite{Untiedt2002, Manohar2012}.
 \label{ltrace_FeD2}}
\end{figure}

In similar experiments for Fe-D$_2$-Fe, the conductance histogram shows a broad peak at 1G$_{0}$ with a small shoulder peak at 0.25G$_{0}$, as illustrated in Fig.~\ref{Gtrace_Ferro}b. The latter feature is enhanced in the logarithmic conductance histogram shown in the inset. In contrast to the Ni-D$_2$-Ni system Fe-D$_2$-Fe shows very long plateaus at lower conductances. Typical traces are shown the inset of Fig.~\ref{ltrace_FeD2}. The Fe-D$_2$-Fe system shows chain lengths of 5 to 6 atoms long quite frequently, and some traces indicate up to 8 atom long chains. The distribution of the lengths of the last plateaus in the breaking traces are collected in the form of length histograms. These are obtained from the lengths of the last plateaus in all traces between suitably chosen start and stop values for the conductance.  In order to capture the last plateau we chose start and stop values of  $1.5\rm{G}_{0}$ and $0.5\rm{G}_{0}$, respectively, based on the position and the width of the peak near $1\rm{G}_{0}$ in the conductance histogram.  We have verified that the resulting histograms do not sensitively depend on this choice. Such length histograms for Ni-D$_2$-Ni junctions do not display any distinct signature of very long chains. On the other hand,  the length histogram of Fe-D$_2$-Fe given in Fig.~\ref{ltrace_FeD2} shows multiple peaks with a spacing between the first two prominent peaks of about 1.63\AA, which indicates the formation of atomic chains of distinct atomic length. 

The first five peaks can loosely be interpreted as counts for 2, 3, 4, 5, 6 atom long chains. The diminishing tail shows the presence of even longer chains, albeit with lower counts. The inter peak distance of $1.63\rm{\AA}$ is close to the experimental value obtained by Thijssen \textit{et al.}~\cite{Thijssen2006} and the value calculated by Bahn \textit{et al.}~\cite{Bahn2002} for the Au-H$_{2}$-Au system. 

\begin{figure}[t!]
\centering
\includegraphics[scale =0.40]{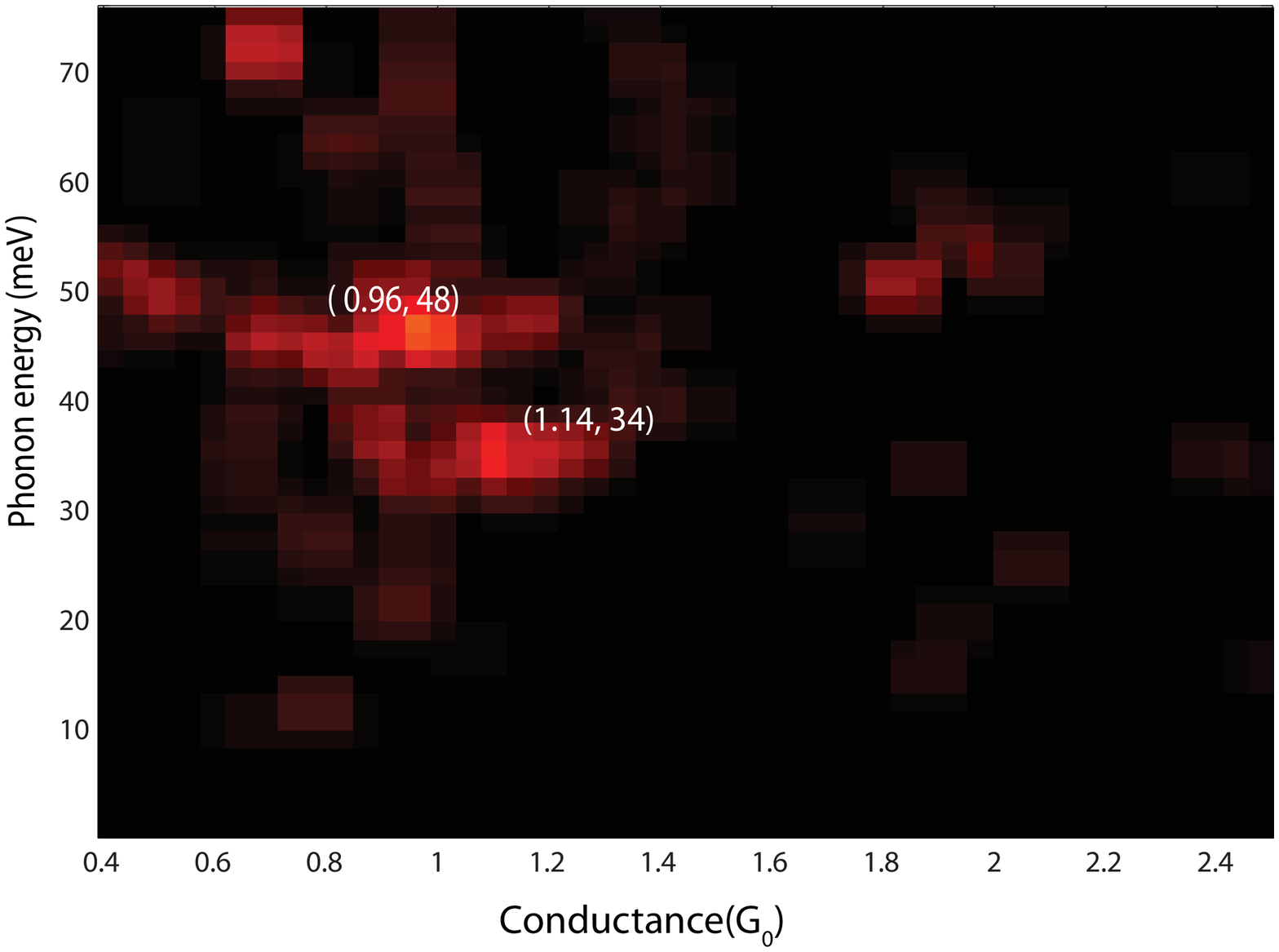}
\caption [Density plot of the energies of the vibrationally induced two level systems (VITLS) for Fe-D$_2$-Fe]{Density plot of the energies of the vibrationally induced two level systems (VITLS) for Fe-D$_2$-Fe ferromagnetic molecular junctions (obtained from the differential conductance for the positive bias) plotted against the zero-bias conductance of the junction. \label{VITLSdens_FerroD2}}
\end{figure}

We proceeded by investigating our molecule assisted ferromagnetic junction by the voltage dependence of the differential conductance (point contact spectroscopy) and shot noise measurements.  The differential conductance is expected to show steps at the energies for vibration mode excitation in the junction. The  sign of the step depends on the transmission probability of the conductance channels~\cite{Galperin2004,Viljas2005,delaVega2006,Tal2008}.  
The differential conductance measurements for the pure ferromagnetic atomic contacts (Fe and Ni) are dominated by the presence of a strong zero-bias anomaly that has been attributed to local Kondo scattering~\cite{Calvo2009}. 
Typical traces for Ni-D$_2$-Ni and Fe-D$_2$-Fe are shown in the supplementary information~\cite{Suppl.Info.}. The differential conductance shows various features: apart from the regular step-down or step-up features anomalous peaks or dips are seen, sometimes along with zero bias anomalies. These anomalous spikes occur at the same energies as the usual vibronic energy steps and have been interpreted in terms of vibrationally induced two level systems (VITLS) in molecular junctions~\cite{Thijssen2006a}. VITLS signals are usually stronger than the regular step structures and can serve as a spectroscopic tool for the study of vibrational energy states in molecular junctions. We observe a large variation in spectroscopic signatures, as may be expected for the many configurations that the atoms can adopt in the junction. The most frequently encountered structures can be read from a density plot of 1000 such differential conductance traces (Fig.~\ref{VITLSdens_FerroD2}), showing VITLS but discarding zero bias anomalies. The plot shows a high density of points at 35meV and 48meV for junctions having a conductance close to 1G$_{0}$. These vibronic energy values are close to values found by Thijssen \textit{et al.} in Pt-D$_2$-Pt atomic chains~\cite{Thijssen2006a}. These VITLS signatures support our conclusion that D$_2$ is part of the structure of the ferromagnetic atomic contact. 

Additionally, we have taken noise spectra on the ferromagnetic molecular junctions. Due to the discreteness of the electronic charge the electronic current fluctuates around its mean value, which is known as shot noise. For a ballistic conductor shot noise is very sensitive to the transmission probabilities of its electronic energy states. An electron wave traversing the junction will be partially transmitted and partially reflected, if the transmission of the particular channel is not perfect. The partial occupation of forward traveling states and backward reflected states give rise to shot noise. Hence, shot noise spectroscopy reveals information on the transmission probabilities and on the numbers of transmission channels involved in electronic transport. For a quantum conductor having $N$ transmission channels with transmission probabilities $\tau_n$ ($0\le \tau_n \le 1$) the conductance $G$ and shot noise $S_I$ can be expressed as \cite{Blanter2000}, 
\begin{equation}
\label{conductance}
G = \frac{e^2}{h}\displaystyle\sum_{n=1}^{N}\tau_i.
\end{equation} 
\begin{equation}
S_{I} = \frac{e^2}{h}\left[2eV\coth(\frac {eV}{2k_{\rm B}T})\displaystyle\sum_{n=1}^{N}\tau_{n}\left(1-\tau_{n}\right) + 4k_{\rm B}T\displaystyle\sum_{i=1}^{N} \tau^2_{n}\right].
\label{shotnoise_1}
\end{equation} 
Here, we have taken the summation over transmission probabilities to include the summation over the spin of the states, {\it i.e.} we sum over spin channels. $V$ is the applied bias voltage and $k_{\rm B}T$ is the thermal energy, with $T$ the bath temperature of the conductor. Eq.~(\ref{shotnoise_1}) includes the Johnson-Nyquist thermal noise, which is the equilibrium noise purely defined by the bath temperature $T$. For $eV \gg k_{\rm B}T$, the non-equilibrium noise dominates and Eq.~(\ref{shotnoise_1}) reduces to $S_I = 2e \langle I\rangle F$, where $\langle I\rangle$ is the time averaged current and $F$ is the Fano factor given by
\begin{equation}
F=\frac{\displaystyle\sum_{n=1}^N \tau_{n}\left(1-\tau_{n}\right)}{\displaystyle\sum_{n=1}^N \tau_{n}}.
\label{Fano_factor}
\end{equation}
 
Shot noise spectroscopy is quite sensitive to local configurations of a quantum conductor. For instance, the shot noise in a coherent Landauer conductor will show sub-Poissonian noise, {\it i.e.} zero -temperature noise below $2e\langle I\rangle$, strictly defined by the transmission probabilities of the transmission channels. On the other hand, the presence of localized states in a conductor will give rise to super-Poissionian noise,  depending on the ratio of traversal and localized time scales of the current carrying electrons~\cite{Belzig2005}. The latter applies to conductors having two-level fluctuators~\cite{Carmi2012}. Hence,  for the purpose of simplifying the analysis we will restrict noise spectroscopy to junctions having no VITLS signatures in the differential conductance, but that only show molecular signatures as step-down or step-up features.  

The experiments started by searching for a stable ferromagnetic molecular junction with chain length of 3 atoms or longer, which was automated using a Labview controlled dc conductance measurement setup. Once a stable contact was found its differential conductance was measured using a lock-in amplifier. Contacts showing only step features in their differential conductance spectra were selected for the shot noise measurement. Noise spectra  were taken at bias settings in steps of 0.25mV up to 3mV, or even 5mV, depending upon the stability of the junction. The molecular junctions are prone to showing $1/f$ noise at higher currents, so that we mostly constraint ourselves to low bias noise measurements and the spectral range above the corner frequency of $1/f$ noise.  At the end of each noise measurement series the thermal noise and differential conductance of the contact was measured again. The differential conductance and thermal noise are sensitive to contact configurations and minuscule changes in the configuration can be detected. Noise spectra were only retained for further analysis when no configurational changes were detected. 

\begin{figure}[h!]
\centering
\includegraphics[scale =0.50]{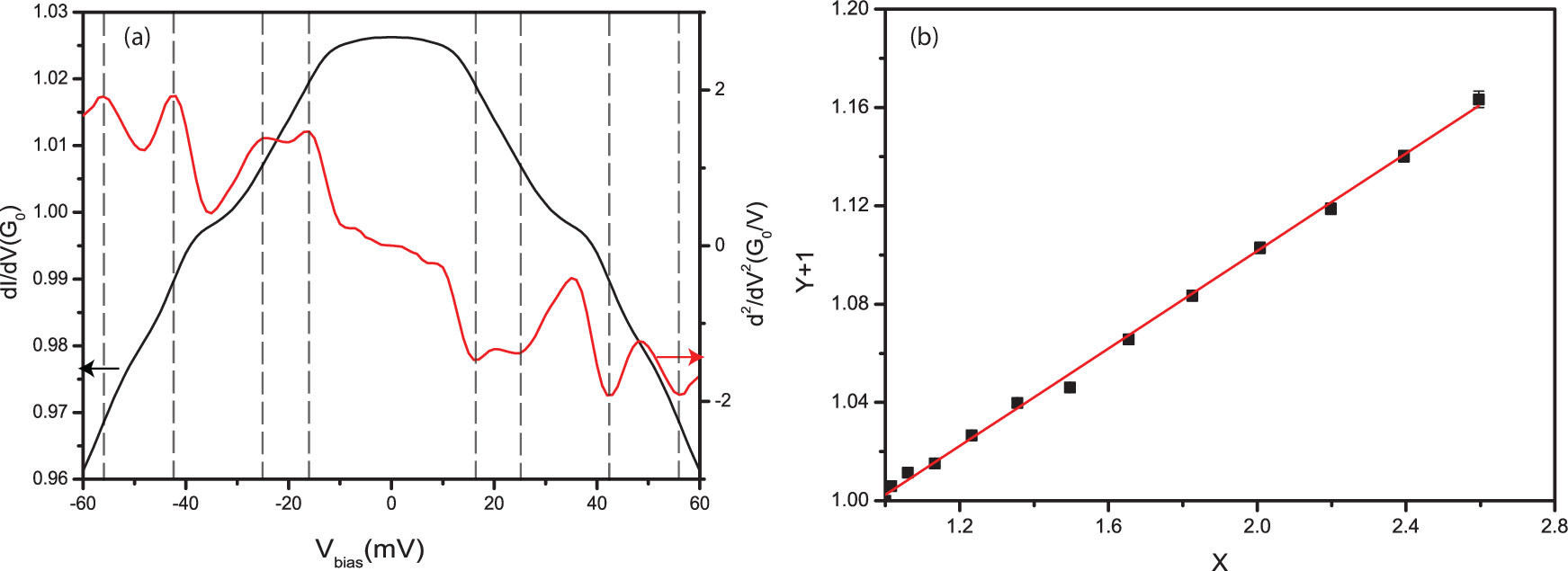}
\caption [Point contact spectrum and shot noise for a Ni-D$_2$-Ni junction.]{Point contact spectrum (a) for a Ni-D$_2$-Ni junction. The inelastic scattering processes are seen in the differential conductance as downward steps, when viewed coming from zero bias. The black curve gives the symmetrized part of the differential conductance; the red curve shows its derivative.
Shot noise data are shown in (b) for the same contact as a reduced axis plot. Here $Y= \frac{S_I(V)-S_I(0)}{S_I(0)} $ and $X = \frac{eV}{2k_{\rm B}T}coth\left[  \frac{2eV}{k_{\rm B}T}\right]$. The Fano factor is obtained from the slope of the linear fit (red line) to the experimental data. The Fano factor for this junction is 0.100$\pm$ 0.002.  \label{didvSN_NiFeD2}} 
\end{figure}

Figure~\ref{didvSN_NiFeD2} shows an example of such measurements for a Ni-D$_2$-Ni junction. In order to suppress contributions from conductance fluctuations (due to electron interference as a result of its scattering on defects in the leads~\cite{Ludoph1999}) the symmetrized differential conductance is shown, $\frac{1}{2} \left[ G(V)+G(-V) \right]$. The differential conductance measurement shows vibronic energy features at 16meV, 25meV, 43meV and 56meV. 
The lowest modes at 16meV and 25meV  may be associated with local vibrations of the Ni atomic contact, or possibly Ni-D states. However, the highest two can only be associated with light elements, pointing at vibronic states associated with a D$_2$ molecule \cite{Djukic2006}. Noise spectra were taken for the very same contact at biasing steps of 0.31mV. Figure~\ref{didvSN_NiFeD2}b shows a plot of the corresponding noise measurement in reduced units~\cite{Manohar2012}. Here, $Y= \left( S_I(V)-S_I(0)\right)/S_I(0) $ is the noise with the thermal noise removed, and scaled to the thermal noise. The change in bias settings can be expressed in the scaled variable
$$
X = \frac{eV}{2k_{\rm B}T}\coth\left[  \frac{2eV}{k_{\rm B}T}\right].
$$ 
Representing noise data in these reduced units simplifies the expression for noise to a linear relation, from which the Fano factor can be directly obtained as
\begin{equation}
F = \frac{Y(V)}{\left( X(V) - 1 \right)}
\label{reduced_axis}
\end{equation}
Thus, the Fano factor can be directly obtained from the slope of the reduced axis plot. 
The Fano factor as obtained from the linear fit to the experimental data is F = 0.100$\pm$ 0.002 . 

\subsection{Discussion and conclusion}

The data in Figure~\ref{didvSN_NiFeD2} permit some conclusions regarding the magnetic state of the conductance channels. The transmission probability of channels are calculated using Eqs.~(\ref{conductance}) and (\ref{Fano_factor}). The small value for the Fano factor implies that all conductance channels either have a small transmission probability, or a transmission close to 1. The combination of measured values for $G/G_{0}=1.03$ and $F=0.100$ are incompatible with less than two spin channels. Expanding the number of channels we find the first solution when assuming three channels. Two of these need to be large, the third small, and a possible solution would be $\tau_{1, \uparrow} = 1.00$, $\tau_{1, \downarrow} = 0.91$ and ($\tau_{2, \uparrow} = 0.15$ or $\tau_{2, \downarrow} = 0.15$).  This solution is not unique, but represents the essential features of all solutions with three channels and corresponds to weakly spin polarized electron transport. When expanding the analysis to include higher numbers of channels the transmission eigenvalues must lie still closer to either 0 or 1, with a maximum of two spin channels having a transmission close to 1. Such solutions are therefore either weakly spin polarized, when the two high transmission channels have opposite spins, or the opposite, strongly spin polarized, when both high-transmission channels have the same spin. The fact that there appears to be one pair of channels close to $\tau =1$ strongly suggests that the s-symmetry derived conductance channels dominate the transmission, and that the weakly spin polarized solution is realized. 
In contrast, the shot noise measurements taken for the pure ferromagnetic atomic contacts show predominantly four, or more, spin channels taking part in the conductance, with no preference for transmission values close to 1. The wide open transmission channels appear to be characteristic for the D$_2$ modified ferromagnetic atomic contacts, in line with earlier observations for Pt-D$_2$-Pt molecular junctions~\cite{Djukic2006}. 

The presence of D$_2$ apparently filters the channels to produce two high-transmission spin channels. Gracia-Saurez \textit{et al.} performed \textit{ab initio} calculations for Pt-H$_2$-Pt and Pd-H$_2$-Pd atomic junctions using the quantum transport code SMEAGOL~\cite{Garcia2005}. They observed that the anti-bonding state of H$_{2}$ hybridizes strongly with the $sd_{{z}^{2}}$ states of the Pt tip atoms, leading to two nearly perfectly transmitting spin channels, while the bonding state is more localized within H$_{2}$ and does not contribute much towards conduction. Direct calculations for ferromagnetic molecular junction are still missing, so that one needs to be cautious in comparing our experimental observations with the model calculations of Gracia-Saurez \textit{et al.}  

Our tentative conclusion of weakly spin polarized transport in Ni-D$_2$ junctions does not imply the absence of local magnetic moments. In our recent study of shot noise related to local magnetic moment formation in Pt atomic chains also found only weakly spin polarized conductance channels. This is still compatible with sizable local magnetic moments, as was  demonstrated by means of density-functional calculations~\cite{Manohar2013}, because the highly spin polarized states in the Pt chains play a minor role in electron transmission.

Very recently two groups succeeded in obtaining evidence for strong spin polarization from shot noise. Burtzlaff {\it et al.}~\cite{BurtzlaffPreprint} used a low-temperature ultra-high vacuum scanning tunneling microscope for the study of single magnetic ad-atoms on top of a Au(111) surface, by means of a Au tip. Touching the ad-atom (Co or Fe) with the tip a junction was formed for which conductance and shot noise was recorded. In this case the Au electrodes serve to filter out a single s-like channel, but the magnetic atom gives rise to a strongly different transmission between the spin directions, and polarization of the transmission channel of up to 60\% was observed. 

Independently, Vardimon {\it et al.}~\cite{VardimonPreprint} performed experiments very similar to the ones presented here, for Ni contacts with addition of oxygen instead of deuterium. They do not obtain evidence for long atomic chains, but oxygen becomes incorporated into the contacts to form Ni-O-Ni junctions. The evidence from shot noise shows very strong spin polarization of the conductance, even up to 100\%. It appears that oxygen plays a very special role, here, with the oxygen p-orbitals coupling strongly to the metal d states, thus filtering out the highly-spin polarized d-orbitals of Ni for participating in electron transport, rather than the weakly spin polarized s-orbitals. 

In conclusion, we find that small atoms or molecules added to ferromagnetic junctions may help to limit the number the number of transmission channels and filter the degree of spin polarization. In addition, in some cases they help to induce spontaneous formation of extended atomic chains forming the junction. Such spin polarized single-channel atomic chains may be a great tool in further experiments on spin polarized quantum transport and spintronics.

\acknowledgments{We are grateful to Zhengpen Baardman for assistance in the experiments and to Bert Crama for technical support. This work is part of the research program of
the Foundation for Fundamental Research on Matter (FOM), which is
financially supported by the Netherlands Organisation for
Scientific Research (NWO). }

\pagebreak
\widetext
\begin{center}
\textbf{\large Supplemental Materials: Molecule-assisted ferromagnetic atomic chain formation}
\end{center}
\setcounter{equation}{0}
\setcounter{figure}{0}
\setcounter{table}{0}
\setcounter{page}{1}
\makeatletter
\renewcommand{\theequation}{S\arabic{equation}}
\renewcommand{\thefigure}{S\arabic{figure}}
\renewcommand{\bibnumfmt}[1]{[S#1]}
\renewcommand{\citenumfont}[1]{S#1}

\section*{Conductance histograms for Ferromagnetic atomic contacts}
\begin{figure}[h!]
\centering
\includegraphics[scale =0.55]{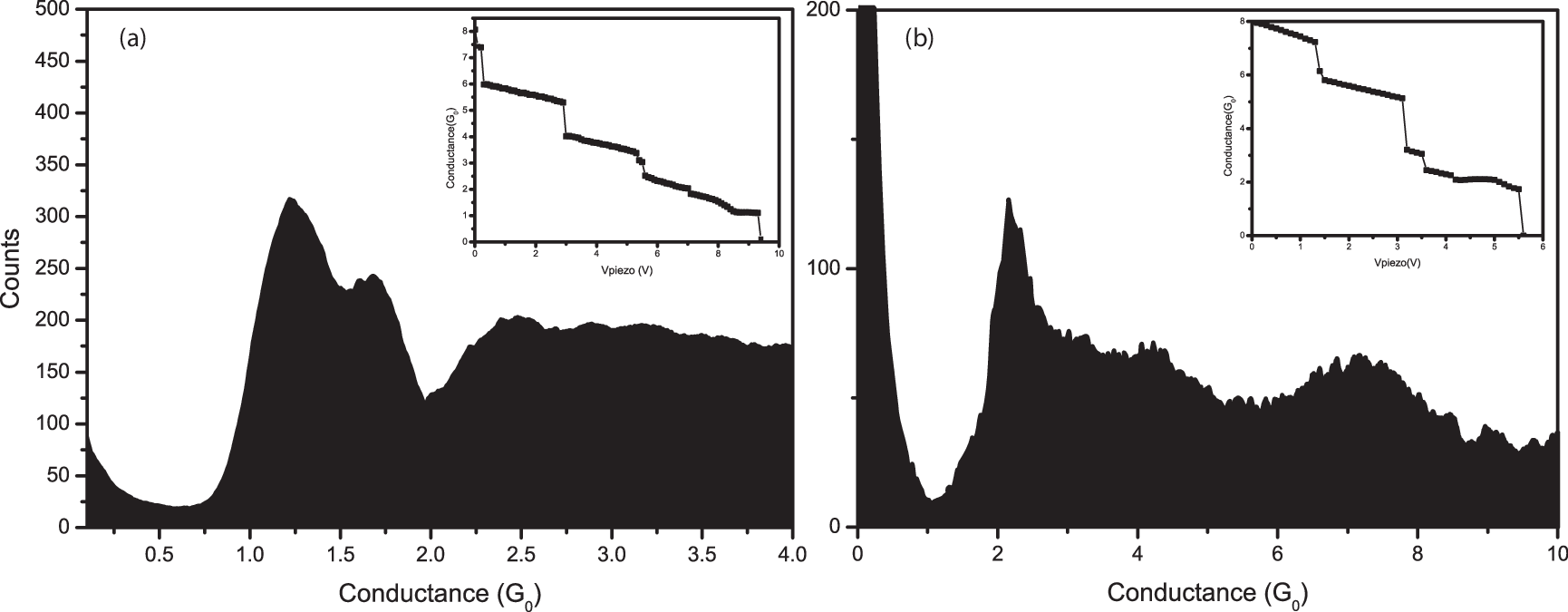}
\caption [Conductance histogram for Ni and Fe atomic contacts]{(a) Conductance histogram for Ni atomic contacts obtained from $1000$ conductance traces, showing peaks at 1.3G$_{0}$ and 1.6G$_{0}$. The inset shows a typical conductance trace. (b) Conductance histogram of Fe atomic contacts obtained from $1000$ conductance traces, showing a peak at 2.1G$_{0}$. Inset shows a conductance trace of Fe atomic contact showing plateau at 2.1G$_{0}$, with conductance monotonically decreasing upon stretching of the contact and eventually breaking of the contact in tunneling regime. \label{GtraceFerro} }
\end{figure}

Ferromagnetic atomic contacts were formed under cryogenic vacuum environment using mechanically controllable break junctions. The ferromagnetic sample wire was first broken using control by a mechanical screw drive and, once the weak link in the wire was broken, further mechanical control was done through a piezo-electric actuator. The contact was reformed by relaxing back the bending of the substrate, and subsequent gently breaking produced atomic-size contacts in the ferromagnetic junction  This ferromagnetic contact is then broken and made repeatedly. The conductance traces were measured at a dc bias voltage of $V=$ 50mV. The conductance traces for Ni show a last plateau at 1.3-1.2G$_{0}$ or 1.6-1.5G$_{0}$ while Fe shows the last plateau around 2G$_{0}$. Several hundred of such traces are collected in form of a conductance histogram. This is shown in figure \ref{GtraceFerro}. The conductance histogram for Ni shows two dominant peaks, one at 1.3G$_{0}$ and other peak at 1.6G$_{0}$ \cite{sCalvo2008}.  The Fe conductance histogram shows a dominant peak at 2.1G$_{0}$ with a drop in counts at 0.7G$_{0}$ and a strong increase in counts in the tunneling regime~\cite{sUntiedt2004}.  
\section*{AC conductance characterization of ferromagnetic molecular junctions}
\begin{figure}[h!]
\centering
\includegraphics[scale =0.55]{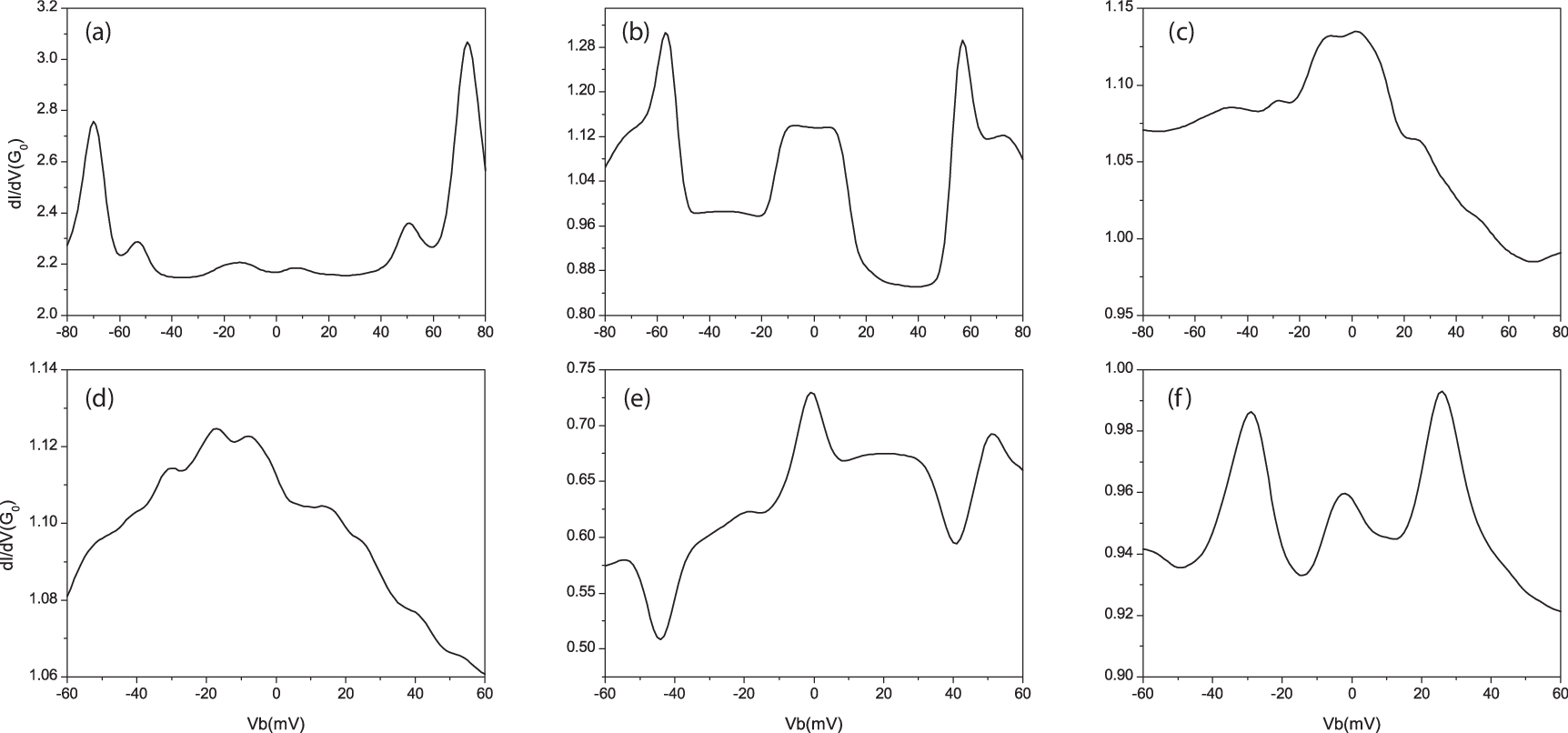}
\caption [Typical differential conductance traces of ferromagnetic Ni-D$_2$ molecular junctions]{Typical differential conductance traces of ferromagnetic Ni-D$_2$ molecular junctions. The vibronic signal associated with D$_2$ can be seen as step-down and vibrational induced two level signals (VITLS, peaks) in the measured differential conductance. The VITLS signature is dominant in most  differential conductance spectra. A zero bias anomaly is also often seen along with VITLS. \label{didv_NiD2} }
\end{figure}

\begin{figure}[h!]
\centering
\includegraphics[scale =0.55]{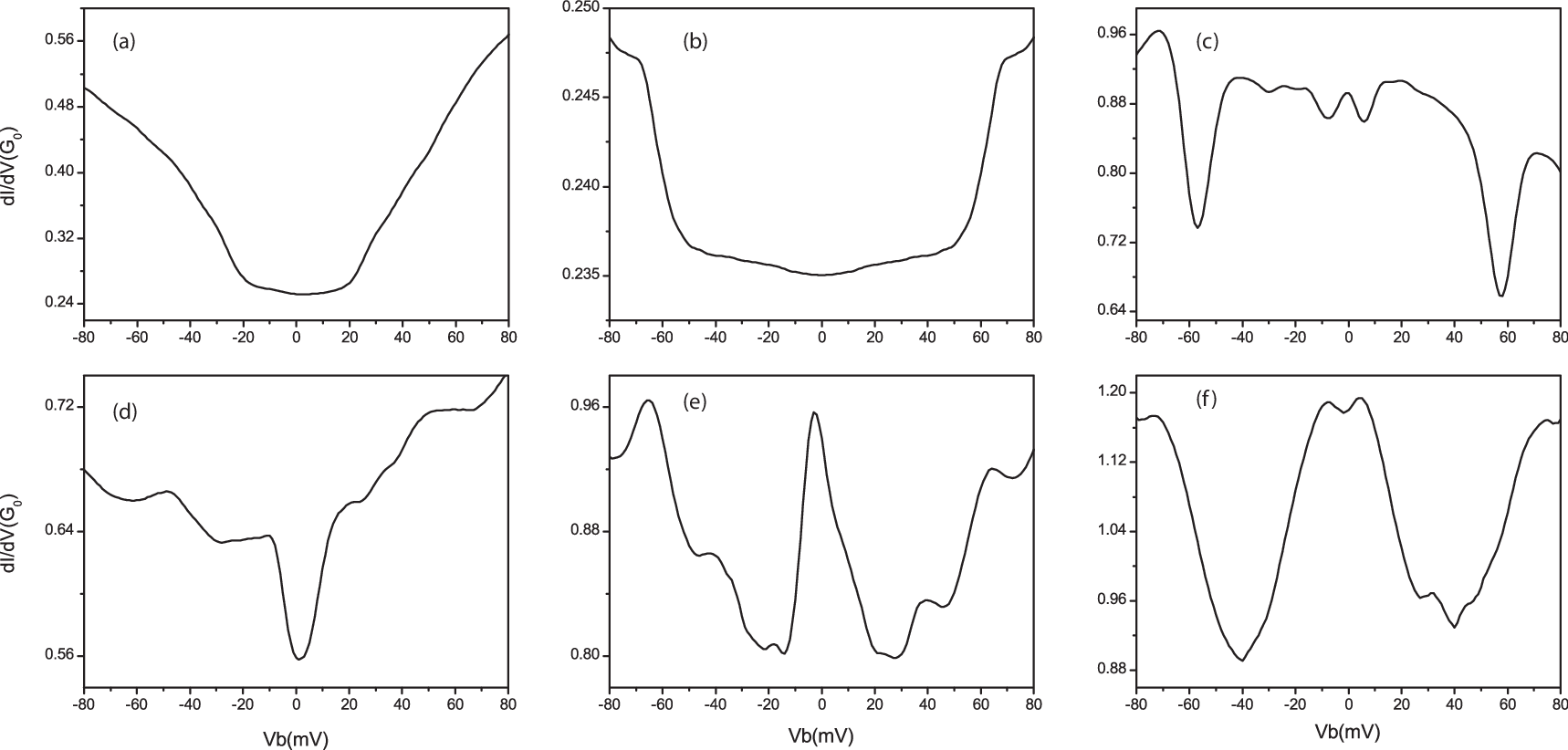}
\caption [Typical differential conductance traces of ferromagnetic Fe-D$_2$ molecular junctions]{Typical differential conductance traces of ferromagnetic Fe-D$_2$ molecular junctions. The vibronic signal associated with D$_2$ can be seen as step-up, step-down and vibrational induced two level signals (VITLS, peaks or dips) in most of the ferromagnetic junctions.  A zero bias anomaly is more frequently prominent in Fe-D$_2$-Fe than in Ni-D$_2$-Ni molecular junctions. \label{didv_FeD2} }
\end{figure}

The ac conductance on ferromagnetic molecular junction was measured using a lockin technique. A sine signal of amplitude 2mV and frequency 2.777kHz is used as reference signal while the dc bias voltage is swept from -80mV to 80mV. The differential conductance measurements obtained in this way for pure ferromagnetic atomic contacts show dominant zero bias anomalies \cite{sCalvo2009}. Addition of D$_2$ partially suppresses this strong anomaly and we observe the appearance of step-up, step-down, or peaks, that we attribute to vibronic energies associated with the presence of D$_2$. Typical differential conductance measurements on ferromagnetic molecular junctions are shown in figures \ref{didv_NiD2} and \ref{didv_FeD2}.


\begin{thebibliography}{99}
\bibitem{Yanson1998}
A. I. Yanson and G. R. Bollinger and H. E. van den Brom and N. Agr\"ait, and J. M. van Ruitenbeek. Formation and manipulation of a metallic wire of single gold atoms.
\newblock {\em Nature}{ \bf 395}, 783 (1998).

\bibitem{Ohnishi1998}
Ohnishi Hideaki and Kondo Yukihito and Takayanagi Kunio. Quantized conductance through individual rows of suspended gold atoms.
\newblock {\em Nature}{ \bf 395}, 780 (1998).

\bibitem{Cuveas1998}
Cuevas, J. C. and Levy Yeyati, A. and Mart\"in-Rodero, A. and Rubio Bollinger, G. and Untiedt, C. and Agr\"ait, N. Evolution of Conducting Channels in Metallic Atomic Contacts under Elastic Deformation.
\newblock {\em Phys. Rev. Lett.}{ \bf 81}, 2990 (1998).

\bibitem{Rubio2003}
Rubio-Bollinger, G. and de las Heras, C. and Bascones, E. and Agr\"ait, N. and Guinea, F. and Vieira, S. Single-channel transmission in gold one-atom contacts and chains.
\newblock {\em Phys. Rev. B.}{ \bf 67}, 121407 (2003).

\bibitem{Scheer2006}
Scheer, E. and Konrad, P. and Bacca, C. and Mayer-Gindner, A. and v. L\"ohneysen, H. and H\"afner, M. and Cuevas, J. C. Correlation between transport properties and atomic configuration of atomic contacts of zinc by low-temperature measurements. 
\newblock {\em Phys. Rev. B.}{ \bf 74}, 205430 (2006).

\bibitem{Bettini2006}
J. Bettini and F. Sato and P. Z. Coura and S. O. Dantas and D. S. Galv\`ao and D. Ugarte. Experimental realization of suspended atomic chains composed of different atomic species. 
\newblock {\em Nature Nanotechnology}{ \bf 1}, 182 (2006).

\bibitem{Thijssen2006}
Thijssen, W. H. A. and Marjenburgh, D. and Bremmer, R. H. and van~Ruitenbeek, J. M. Oxygen-Enhanced Atomic Chain Formation.
\newblock {\em Phys. Rev. Lett.}{ \bf 96}, 026806 (2006).

\bibitem{Csonka2006}
Csonka, S. and Halbritter, A. and Mih\'aly, G. Pulling gold nanowires with a hydrogen clamp: Strong interactions of hydrogen molecules with gold nanojunctions. 
\newblock {\em Phys. Rev. B.}{ \bf 73}, 075405 (2006).

\bibitem{Fischer2007}
Fischer, Marinus and van Houselt, Arie and Kockmann, Daan and Poelsema, Bene and Zandvliet, Harold J. W.  Pulling gold nanowires with a hydrogen clamp: Formation of atomic Pt chains on Ge(001) studied by scanning tunneling microscopy.
\newblock {\em Phys. Rev. B.}{ \bf 76}, 245429 (2007).

\bibitem{Halbriter2007}
Andr\'as Halbritter and Szabolcs Csonka and P\'eter Makk and Gy\"orgy Mih\'aly. Interaction of hydrogen with metallic nanojunctions.
\newblock {\em Journal of Physics: Conference serries }{ \bf 61}, 214 (2007).

\bibitem{Smit2009}
R. H. M. Smit and A I Mares and M H\"afner and P Pou and J C Cuevas and J M van Ruitenbeek. Metallic properties of magnesium point contacts. 
\newblock {\em New Journal of Physics}{ \bf 11}, 073043 (2009).

\bibitem{Makk2012}
P. Makk and Z. Balogh and S. Csonka and A. Halbritter.  Pulling platinum atomic chains by carbon monoxide molecules. 
\newblock {\em Nanoscale}{ \bf 4}, 4739 (2012).

\bibitem{Untiedt2004}
Untiedt, C. and Dekker, D. M. T. and Djukic, D. and van Ruitenbeek, J. M.  Absence of magnetically induced fractional quantization in atomic contacts. 
\newblock {\em  Phys. Rev. B.}{ \bf 69}, 081401 (2004).

\bibitem{Calvo2008}
M. R. Calvo and M. J. Caturla and D. Jacob, C and Untiedt and J. J. Palacios. Mechanical, Electrical, and Magnetic Properties of Ni Nanocontacts. 
\newblock {\em IEEE Transaction on Nanotechnology}{ \bf 7}, 165 (2008).

\bibitem{Calvo2009}
M. R. Calvo and Joaqu\'in Fern\'andez-Rossier and Juan Jos\'e Palacios and David Jacob and Douglas Natelson and Carlos Untiedt.  The Kondo effect in ferromagnetic atomic contacts.
\newblock {\em  Nature}{ \bf 458}, 1150 (2009).

\bibitem{Nakazumi2010}
Nakazumi, T. and Kiguchi, M. Formation of Co Atomic Wire in Hydrogen Atmosphere. 
\newblock {\em J. Phys. Chem. Lett.}{ \bf 1}, 923 (2010).

\bibitem{Kiguchi2010}
Kiguchi, Manabu and Hashimoto, Kunio and Ono, Yuriko and Taketsugu, Tetsuya and Murakoshi, Kei. Formation of a Pd atomic chain in a hydrogen atmosphere. 
\newblock {\em Phys. Rev. B.}{ \bf 81}, 195401 (2010).

\bibitem{Suppl.Info.}
Manohar Kumar and Kiran Kumar Vidya Sethu and J. M. van Ruitenbeek. Molecule-assisted ferromagnetic atomic chain formation: Supplementary information.
\newblock {\em}{\bf }(2014).

\bibitem{Smit2002}
R. H. M. Smit and Y. Noat and C. Untiedt and N. D. Lang and M. C. van Hemert and J. M. van Ruitenbeek. Measurement of the conductance of a hydrogen molecule. 
\newblock {\em Nature}{ \bf 419 }, 906 (2002).

\bibitem{Manoharthesis2012}
Manohar Kumar.  A study of electron scattering through noise spectroscopy.
\newblock {\em Leiden University}{\bf  } (2012).

\bibitem{Brom1999}
van den Brom, H. E. and van Ruitenbeek, J. M.  Quantum Suppression of Shot Noise in Atom-Size Metallic Contacts.
\newblock {\em Phys. Rev. Lett.}{ \bf 82}, 1526 (1999).

\bibitem{Djukic2006}
D. Djukic and J. M. van Ruitenbeek.  Shot Noise Measurements on a Single Molecule.
\newblock {\em Nano Letters}{ \bf 6}, 789 (2006).

\bibitem{Kumar1996}
Kumar, A. and Saminadayar, L. and Glattli, D. C. and Jin, Y. and Etienne, B.  Experimental Test of the Quantum Shot Noise Reduction Theory. 
\newblock {\em Phys. Rev. Lett.}{ \bf 76}, 2778 (1996).

\bibitem{Untiedt2002}
Untiedt, C. and Yanson, A. I. and Grande, R. and Rubio-Bollinger, G. and Agr\"ait, N. and Vieira, S. and van Ruitenbeek, J.M.  Calibration of the length of a chain of single gold atoms.
\newblock {\em Phys. Rev. B.}{ \bf 66}, 085418 (2002).

\bibitem{Manohar2012}
Kumar, Manohar and Avriller, R\'emi and Yeyati, Alfredo Levy and van~Ruitenbeek, Jan M.  Detection of Vibration-Mode Scattering in Electronic Shot Noise.
\newblock {\em Phys. Rev. Lett.}{ \bf 108}, 146602 (2012).

\bibitem{Bahn2002}
Bahn, S. R. and Lopez, N. and N\o{}rskov, J. K. and Jacobsen, K. W.  Adsorption-induced restructuring of gold nanochains.
\newblock {\em Phys. Rev. B.}{ \bf 66}, 081405 (2002).

\bibitem{Galperin2004}
M. Galperin, and M. A. Ratner, and A. Nitzan.  Inelastic electron tunneling spectroscopy in molecular junctions: Peaks and dips. 
\newblock {\em J. Chem. Phys.}{ \bf 121}, 11965 (2004).

\bibitem{Viljas2005}
J. K. Viljas, and J. C. Cuevas, and F. Pauly, and M. H\"afner. Electron-vibration interaction in transport through atomic gold wires.
\newblock {\em Phys. Rev. B.}{ \bf 72}, 245415 (2005).

\bibitem{delaVega2006}
L. de la Vega, and A. Mart´ın-Rodero, and N. Agr\"aıt, and A. Levy Yeyati. Universal features of electron-phonon interactions in atomic wires.
\newblock {\em Phys. Rev. B.}{ \bf 73}, 075428 (2006).

\bibitem{Tal2008}
Tal, O. and Krieger, M. and Leerink, B. and van Ruitenbeek, J. M. Electron-Vibration Interaction in Single-Molecule Junctions: From Contact to Tunneling Regimes.
\newblock {\em Phys. Rev. Lett.}{ \bf 100}, 196804 (2008).

\bibitem{Thijssen2006a}
Thijssen, W. H. A. and Djukic, D. and Otte, A. F. and Bremmer, R. H. and van Ruitenbeek, J. M.  Vibrationally Induced Two-Level Systems in Single-Molecule Junctions. 
\newblock {\em Phys. Rev. Lett.}{ \bf 97}, 226806 (2006).

\bibitem{Blanter2000}
Ya.M. Blanter and M. B\"uttiker. Shot noise in mesoscopic conductors. 
\newblock {\em Physics Reports}{ \bf 336}, 1 (2000).

\bibitem{Belzig2005}
Belzig, W. Full counting statistics of super-Poissonian shot noise in multilevel quantum dots.  
\newblock {\em Phys. Rev. B.}{ \bf 71}, 161301 (2005).

\bibitem{Carmi2012}
Carmi, Assaf and Oreg, Yuval. Enhanced shot noise in asymmetric interacting two-level systems. 
\newblock {\em Phys. Rev. B.}{ \bf 85}, 045325 (2012).

\bibitem{Ludoph1999}
B. Ludoph, and M. H. Devoret, and D. Esteve, and C. Urbina, and J. M. van Ruitenbeek. Evidence for Saturation of Channel Transmission from Conductance Fluctuations in Atomic-Size Point Contacts.
\newblock {\em Phys. Rev. Lett.}{ \bf 82}, 1530 (1999).

\bibitem{Garcia2005}
Garc\'ia-Su\'arez, V. M. and Rocha, A. R. and Bailey, S. W. and Lambert, C. J. and Sanvito, S. and Ferrer, J. Single-channel conductance of ${\mathrm{H}}_{2}$ molecules attached to platinum or palladium electrodes. 
\newblock {\em Phys. Rev. B.}{ \bf 72}, 045437 (2005).

\bibitem{Manohar2013}
Kumar, Manohar and Tal, Oren and Smit, Roel H. M. and Smogunov, Alexander and Tosatti, Erio and van Ruitenbeek, Jan M. Shot noise and magnetism of Pt atomic chains: Accumulation of points at the boundary. 
\newblock {\em Phys. Rev. B.}{ \bf 88}, 245431 (2013).

\bibitem{BurtzlaffPreprint}
Burtzlaff, A. and Weismann, A. and Brandbyge, M. and Berndt, R.Shot noise as a probe of spin polarized transport through single atoms. 
\newblock {\em Preprint}{ \bf } (2014).

\bibitem{VardimonPreprint}
Vardimon, R. and Klionsky, M. and Tal, O. Fully spin-polarized conductance in atomic-scale junctions. 
\newblock {\em Preprint}{ \bf } (2014).


\end{thebibliography}

\begin{thebibliography}{9}

\bibitem{sCalvo2008}
M. R. Calvo and M. J. Caturla and D. Jacob, C and Untiedt and J. J. Palacios. Mechanical, Electrical, and Magnetic Properties of Ni Nanocontacts.
\newblock {\em IEEE Transaction on Nanotechnology}{ \bf 7}, 165--168 (2008).

\bibitem{sUntiedt2004}
C. Untiedt and D. M. T. Dekker and D. Djukic, and J. M. van Ruitenbeek. Absence of magnetically induced fractional quantization in atomic contacts.
\newblock {\em Physical Rev. B.}{ \bf 69}, 081401(R) (2004).

\bibitem{sCalvo2009}
M. Reyes Calvo and Joaqu\'in Fern\'andez-Rossier and Juan Jos\'e
	Palacios and David Jacob and Douglas Natelson and Carlos Untiedt. The Kondo effect in ferromagnetic atomic contacts
\newblock {\em Nature}{ \bf 458}, 1150--1153 (2009).



\end{thebibliography}
\end{document}